\documentclass[
 reprint,
 superscriptaddress,
 amsmath,amssymb,
]{revtex4-2}

\usepackage{graphicx}% Include figure files
\usepackage{dcolumn}% Align table columns on decimal point
\usepackage{bm}% bold math
\usepackage{color}
\usepackage{xcolor}
\usepackage{times}
\usepackage[hidelinks]{hyperref}
\usepackage{braket}
\usepackage{siunitx}

\hypersetup{
  colorlinks   = true, %Colours links instead of ugly boxes
  urlcolor     = blue, %Colour for external hyperlinks
  linkcolor    = blue, %Colour of internal links
  citecolor   = red %Colour of citations
}

\newcommand{\Yb}{\ensuremath{^{171}\mathrm{Yb}^+~}}

\begin{document}
%\preprint{APS/123-QED}

%\title{Simulation of frustrated quantum magnets in two-dimensional ion crystals}
\title{Observing frustrated quantum magnetism in two-dimensional ion crystals}

 \affiliation{
State Key Laboratory of Low Dimensional Quantum Physics, Department of Physics, Tsinghua University, Beijing 100084, China
 }
 \affiliation{
 Beijing Academy of Quantum Information Sciences, Beijing 100193, China
 }
 \affiliation{
 Frontier Science Center for Quantum Information, Beijing 100084, People’s Republic of China
 }
 \affiliation{
 Current Address: Duke Quantum Center, Duke University, Durham, NC 27708, USA
 }

\author{Mu Qiao}
 \email{
 mu.q.phys@gmail.com
 }
 \affiliation{
State Key Laboratory of Low Dimensional Quantum Physics, Department of Physics, Tsinghua University, Beijing 100084, China
 }

\author{Zhengyang Cai}
 \affiliation{
State Key Laboratory of Low Dimensional Quantum Physics, Department of Physics, Tsinghua University, Beijing 100084, China
 }

\author{Ye Wang}
 \affiliation{
State Key Laboratory of Low Dimensional Quantum Physics, Department of Physics, Tsinghua University, Beijing 100084, China
 }
 \affiliation{
 Current Address: Duke Quantum Center, Duke University, Durham, NC 27708, USA
 }

\author{Botao Du}
 \affiliation{
State Key Laboratory of Low Dimensional Quantum Physics, Department of Physics, Tsinghua University, Beijing 100084, China
 }

\author{Naijun Jin}
 \affiliation{
State Key Laboratory of Low Dimensional Quantum Physics, Department of Physics, Tsinghua University, Beijing 100084, China
 }

\author{Wentao Chen}
 \affiliation{
State Key Laboratory of Low Dimensional Quantum Physics, Department of Physics, Tsinghua University, Beijing 100084, China
 }

\author{Pengfei Wang}
 \affiliation{
State Key Laboratory of Low Dimensional Quantum Physics, Department of Physics, Tsinghua University, Beijing 100084, China
 }
 \affiliation{
 Beijing Academy of Quantum Information Sciences, Beijing 100193, China
 }

\author{Chunyang Luan}
 \affiliation{
State Key Laboratory of Low Dimensional Quantum Physics, Department of Physics, Tsinghua University, Beijing 100084, China
 }

\author{Erfu Gao}
 \affiliation{
State Key Laboratory of Low Dimensional Quantum Physics, Department of Physics, Tsinghua University, Beijing 100084, China
 }

\author{Ximo Sun}
 \affiliation{
State Key Laboratory of Low Dimensional Quantum Physics, Department of Physics, Tsinghua University, Beijing 100084, China
 }

\author{Haonan Tian}
 \affiliation{
State Key Laboratory of Low Dimensional Quantum Physics, Department of Physics, Tsinghua University, Beijing 100084, China
 }

\author{Jingning Zhang}
 \affiliation{
 Beijing Academy of Quantum Information Sciences, Beijing 100193, China
 }

\author{Kihwan Kim}
 \email{
 kimkihwan@mail.tsinghua.edu.cn
 }
 \affiliation{
State Key Laboratory of Low Dimensional Quantum Physics, Department of Physics, Tsinghua University, Beijing 100084, China
 }
 \affiliation{
 Beijing Academy of Quantum Information Sciences, Beijing 100193, China
 }
 \affiliation{
 Frontier Science Center for Quantum Information, Beijing 100084, People’s Republic of China
 }

\date{\today}
\begin{abstract}
Two-dimensional (2D) quantum magnetism is a paradigm in strongly correlated many-body physics \cite{diep2013frustrated,schmidt2017frustrated}. The understanding of 2D quantum magnetism can be expedited by employing a controllable quantum simulator that faithfully maps 2D-spin Hamiltonians. The 2D quantum simulators can exhibit exotic phenomena such as frustrated quantum magnetism, topological order \cite{diep2013frustrated,schmidt2017frustrated,moessner_2006_geometrical_frustration,Balents_2010_frustration,qi_2011_topological,wen_topological_phases,spin_liquid_rmp,Broholm_2020_Spin_Liquid} and can be used to show quantum computational advantages \cite{2d_ising_fermion,spin_ice_rmp, bermejo2018architectures,Qiu2020Programmable}. Many experimental platforms are being developed, including Rydberg atoms and superconducting annealers \cite{dwave_2021_ice,Ebadi_2021_quantum,Scholl_2021_quantum}. However, with trapped-ion systems, which showed the most advanced controllability\cite{Kim_2010_quantum,Rajibul_2013_emergence,Monroe2021Programmable} and quantum coherence \cite{Wang_2017_10minutes,wang_2021_single}, quantum magnetism was explored in one-dimensional chains. Here, we report simulations of frustrated quantum magnetism with 2D ion crystals. We create a variety of spin-spin interactions for quantum magnets, including those that exhibit frustration by driving different vibrational modes and adiabatically prepare the corresponding ground states. The experimentally measured ground states are consistent with the theoretical predictions and are highly degenerate for geometrically frustrated spin models in two dimensions. Quantum coherence of the ground states is probed by reversing the time evolution of the B-field to the initial value and then measuring the extent to which the remaining state coincides with the initial state. Our results open the door for quantum simulations with 2D ion crystals.

\end{abstract}

\maketitle

Developing controllable two-dimensional (2D) systems that can be described by quantum spin models is of great importance. 2D systems in which spin-spin interactions can be programmed can exhibit the exotic phenomena expected in 2D quantum magnets, such as spin frustration \cite{moessner_2006_geometrical_frustration,Balents_2010_frustration}, topological phase \cite{qi_2011_topological,wen_topological_phases}, and spin-liquid phase \cite{spin_liquid_rmp,Broholm_2020_Spin_Liquid}. Quantum simulations performed with these 2D systems allow the study of ground states, excitations, and dynamics of 2D quantum magnets beyond the limits of conventional computational power. The properties of magnetic materials are determined by the intrinsic interactions, such as ferromagnetic or anti-ferromagnetic interactions, between the arranged atoms. Recently, unexpected long-range ferromagnetic orders have been observed in 2D magnetic atomic crystals \cite{Gong_2017_ferromagnetic_2d,Huang_2017_2d_ferromagnets}, and the coexistence of antiferromagnetic and ferromagnetic interactions have been reported to produce non-trivial magnetic ground states \cite{song_2021_moire}. However, creating 2D quantum systems with controllable spin-spin interactions and characterizing the quantum ground states of quantum magnets remain challenging.

Various physical platforms are being developed to simulate 2D quantum magnetism. 2D spin ices have been created using superconducting annealers \cite{dwave_2021_ice}; however, limited coherence prevents the probing of quantum properties. Large-scale quantum magnets have been demonstrated using neutral atoms in 2D lattices \cite{Ebadi_2021_quantum,Scholl_2021_quantum}; however, the blockade interaction precludes the simulation of quantum magnets beyond antiferromagnetism \cite{Qiu2020Programmable}. As a leading platform for quantum simulation, trapped atomic ions exhibit exceptional coherence time \cite{ Wang_2017_10minutes,wang_2021_single}, and can realize both ferromagnetic and anti-ferromagnetic interactions \cite{Kim_2010_quantum,Rajibul_2013_emergence,Monroe2021Programmable}. Previous studies on simulating quantum magnetism were limited to a 1D chain of ions, which is difficult to readily simulate the 2D spin models \cite{bermudez2011frustrated,nath2015hexagonal}. Controllable spin-spin interactions have been shown using 2D-ion crystals in a Penning trap \cite{Britton_2012_engineered,Bollinger_spin_dynamics}. However, the simulation of quantum magnets and the detection of their ground states using 2D ion crystals has not yet been realized.

The 2D ion crystals suffer from the micromotion synchronized with the RF field in the Paul trap, which undermines the quality of quantum simulations. This problem can be overcome by using traps without excess micromotion, like arrays of micro Paul traps in which each trap contains only one micromotion-free ion \cite{Sterling_2014,Mielenz_2016,hakelberg2019interference} or Penning traps with no oscillating fields \cite{Britton_2012_engineered,safavi_2018_verification}. Alternatively, it was pointed out that the detrimental effects of micromotion on quantum simulation can be mitigated by using an oblate Paul trap \cite{Yoshimura_2015,richerme2016two}. Similarly, we developed a monolithic trap that can suppress the influence of micromotion by setting the net-propagation direction of the operation lasers perpendicular to micromotion \cite{wang_2020_coherently}.  

Here, we simulate quantum magnets and directly observe their ground states using 2D ion crystals confined in a monolithic Paul trap. In our quantum simulation, various Ising interactions including ferromagnetic and anti-ferromagnetic ones are generated, and their ground states are prepared adiabatically. With geometrically competing interactions, we observe that the ground states of the simulated magnetic model are frustrated and degenerate. We probe the coherence of quantum simulation, by reversing the adiabatic evolution and measuring how much the remaining state coincides with the initial state.

%\section{Static 2D ion crystals in a Paul trap}

\begin{figure*}[!htb]
\center{\includegraphics[width=\textwidth]{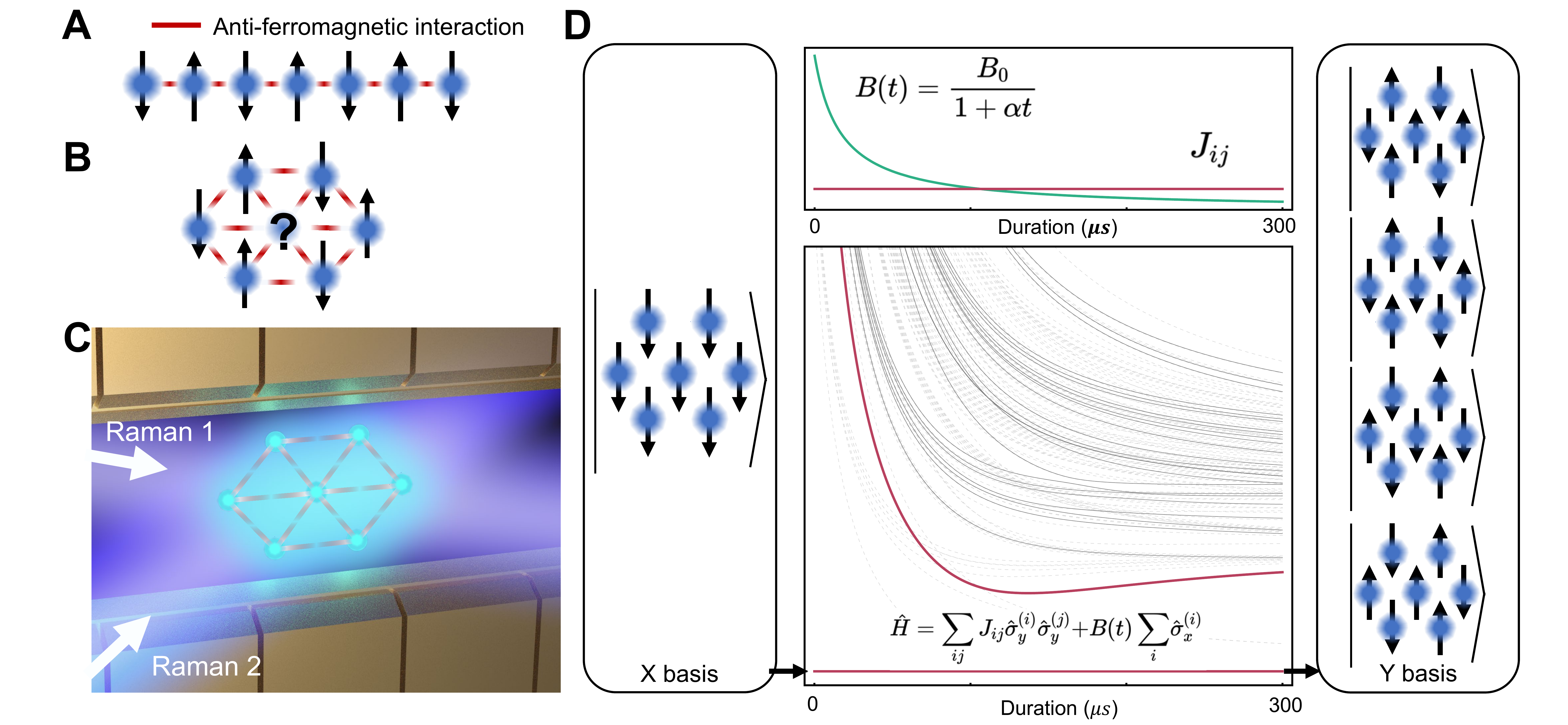}}
\caption{\label{fig:setup} {\bf Quantum simulation of frustrated quantum magnets with 2D ion crystal.} {\bf a}, 1D anti-ferromagnetic quantum magnets present Néel-like ground states, with an example of seven spins. {\bf b}, 2D anti-ferromagnetic quantum magnets exhibit geometric frustration in the ground state. {\bf c}, Atomic ions are trapped by a monolithic trap and controlled by globally illustrated Raman lasers. 
{\bf d}, Systematic diagram of adiabatic quantum simulation for the case of geometrically frustrated 2D spins. The experiment starts from the ground state of the transverse B field, $B\sum_{i}\sigma_{x}^{(i)}$, and adiabatically evolves into the ground state of the Ising Hamiltonian, $\sum_{i,j}J_{i,j}\sigma_{y}^{(i)}\sigma_{y}^{(j)}$, in the case of Fig.1{\bf b}. In the energy-level diagram, the vertical axis represents the energy difference from the ground state. The solid lines represent excited states coupled to the ground state, and the dashed lines represent the other excited states. The red lines represent the ground state and the lowest excited energy level coupled to the ground state. The energy level is scaled with the spin-spin interaction, $J_{0}$. Here, we ramp down the strength of the transverse B field with the profile, $1/(1+\alpha t)$, where $\alpha$ denotes a tuning parameter, and the spin-spin interaction remains constant during the ramping. Lastly, the system reaches the ground state of the frustrated magnet, which is a superposition of four degenerate states owing to competing interactions.
}
\end{figure*}

The fundamental difference between 1D and 2D quantum magnets can be illustrated with the anti-ferromagnetic Ising interaction, which prefers to align neighboring spins in opposite directions. For a 1D quantum system consisting of seven spins with nearest-neighbor anti-ferromagnetic interaction, the ground state is a Néel state with alternating spin directions, as shown in Fig.1{\bf a}. However, if the seven spins are arranged in the 2D triangular lattice as shown in Fig.1{\bf b}, the center spin experiences competition being in the $\ket{\uparrow}$ and $\ket{\downarrow}$ states since three of the neighboring spins push it into the $\ket{\uparrow}$ state while the other three push it into the $\ket{\uparrow}$ state. The competition between different configurations is the characteristic of frustration, which leads to large ground state degeneracy \cite{moessner_2006_geometrical_frustration}.  

\begin{figure*}[!htb]
\center{\includegraphics[width=0.6\textwidth]{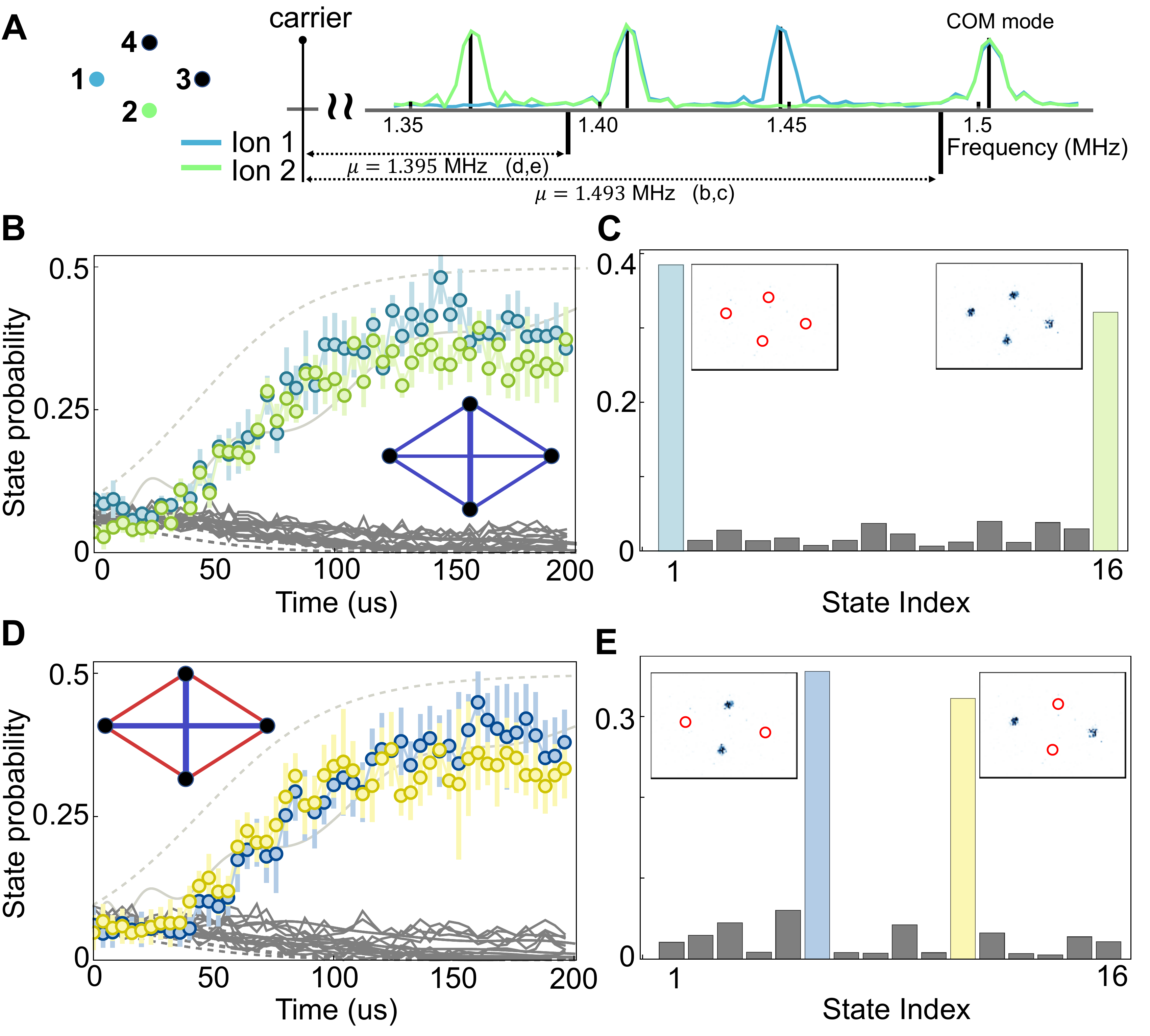}}
\caption{\label{fig:4-ion}{{\bf Quantum simulation with a four-ion 2D crystal.} {\bf a}, Vibrational spectrum of the crystal. The red and green curves represent the fluorescence of red and green ions, respectively. The vertical red lines indicate the mode frequencies, while the solid line represents the detuning used in the experiment. {\bf b},{\bf d} depict the time evolution of different Ising interactions at the detunings of Raman laser beams shown in {\bf a}. Insets present the interaction diagram, where the red and blue lines represent the anti-ferromagnetic and ferromagnetic interactions respectively. 
The filled circles, solid curves, and dashed curves represent the experimental data, theoretical evolution expected from the actual ramp, and populations in the exact ground state, respectively. {\bf c},{\bf e}, depict the experimentally measured populations of the state in binary order at the end of the ramp, which indicates the ground states of the Hamiltonians with the given interaction diagrams. Insets represent reconstructed images based on the binary detection of spin states. {\bf b},{\bf c} Experiment results for the all-to-all ferromagnetic interaction. {\bf d},{\bf e} Experiment results for the nearest anti-ferromagnetic interaction and next-nearest ferromagnetic interaction. Here, we use a transverse B field with a strength of $(2\pi)\times$\qty{29}{\kilo\hertz}, and Raman beams with a strength of $(2\pi)\times$\qty{50}{\kilo\hertz} to generate Ising interactions. The error bars represent standard deviation.}
}
\end{figure*}

Our experiments are performed with \Yb ions in a monolithic Paul trap. The $\ket{F=1,m=0}$ and the $\ket{F=0,m=0}$ states in the $S_{1/2}$ manifold with the energy splitting, $\omega_{\rm HF}=$ \qty{12.642812}{\giga\hertz}, represent the $\ket{\uparrow}$ and $\ket{\downarrow}$ states of a spin-$1/2$ system, which are first-order insensitive to magnetic noise and have a coherence time of over 1 hour \cite{wang_2021_single, Wang_2017_10minutes}. As shown in Fig.1{\bf c}, a 2D ion crystal is confined in a monolithic Paul trap with spin-spin interactions mediated by collective phonon modes through Raman excitations. The electrodes of the trap were designed to rotate the crystal plane along the direction of micromotion. The detrimental effect of micromotion on quantum simulation is thus mitigated by making the net-propagation vector of the Raman laser beams perpendicular to the plane of the crystal. 

Our protocol of adiabatic quantum simulation is shown in Fig.1{\bf d}. The adiabatic evolution starts from the ground state of a simple Hamiltonian, i.e., a transverse $B$-field and the system adiabatically evolves into the ground states of the Ising models by slowly decreasing the $B$-field. The results of the adiabatic quantum simulation are directly measured, which reveals the ground state of the simulated Ising model. 

By coupling ions to Raman laser beams, we realize the Hamiltonian of the transverse field Ising model, 
\begin{equation}
\hat{H}=\sum_{i,j}J_{i,j}\hat{\sigma}_y^{(i)}\hat{\sigma}_y^{(j)}+B(t)\sum_{i}\hat{\sigma}_x^{(i)},
\end{equation} 
where $J_{ij}$ represents the interaction strength between the $i$-th and the $j$-th spins, and $B(t)$ represents the strength of the transverse $B$-field. We generate the transverse $B$-field using Raman laser beams with frequency difference of $\omega_{\rm{HF}}$. To generate the spin-spin interactions, we use bichromatic Raman beams with frequency differences of $\omega_{\rm{HF}}\pm(\nu-\delta)$. The effective Ising interaction is expressed as \cite{cirac_interaction_iontrap, cirac_spin_phases, kihwan_interaction_iontrap}: 
\begin{equation}
    J_{ij}=\Omega_{i}\Omega_{j}\frac{\hbar(\delta k)^2}{2M}\sum_m\frac{b_{i,m}b_{j,m}}{\mu^2-\omega_m^2}
\end{equation}
where $M$ is the mass of the \Yb ion, $\Omega_i$ is the laser Rabi frequency on $i$-th ion, $\delta k$ is the net-propagation vector of the Raman beams, $b_{i,m}$ is the normal mode vector, $\omega_{m}$ is the $m$-th mode frequency, and $\mu$ is the Raman detuning from the $\omega_{\rm HF}$.

\begin{figure*}[!htb]
\center{\includegraphics[width=0.9\textwidth]{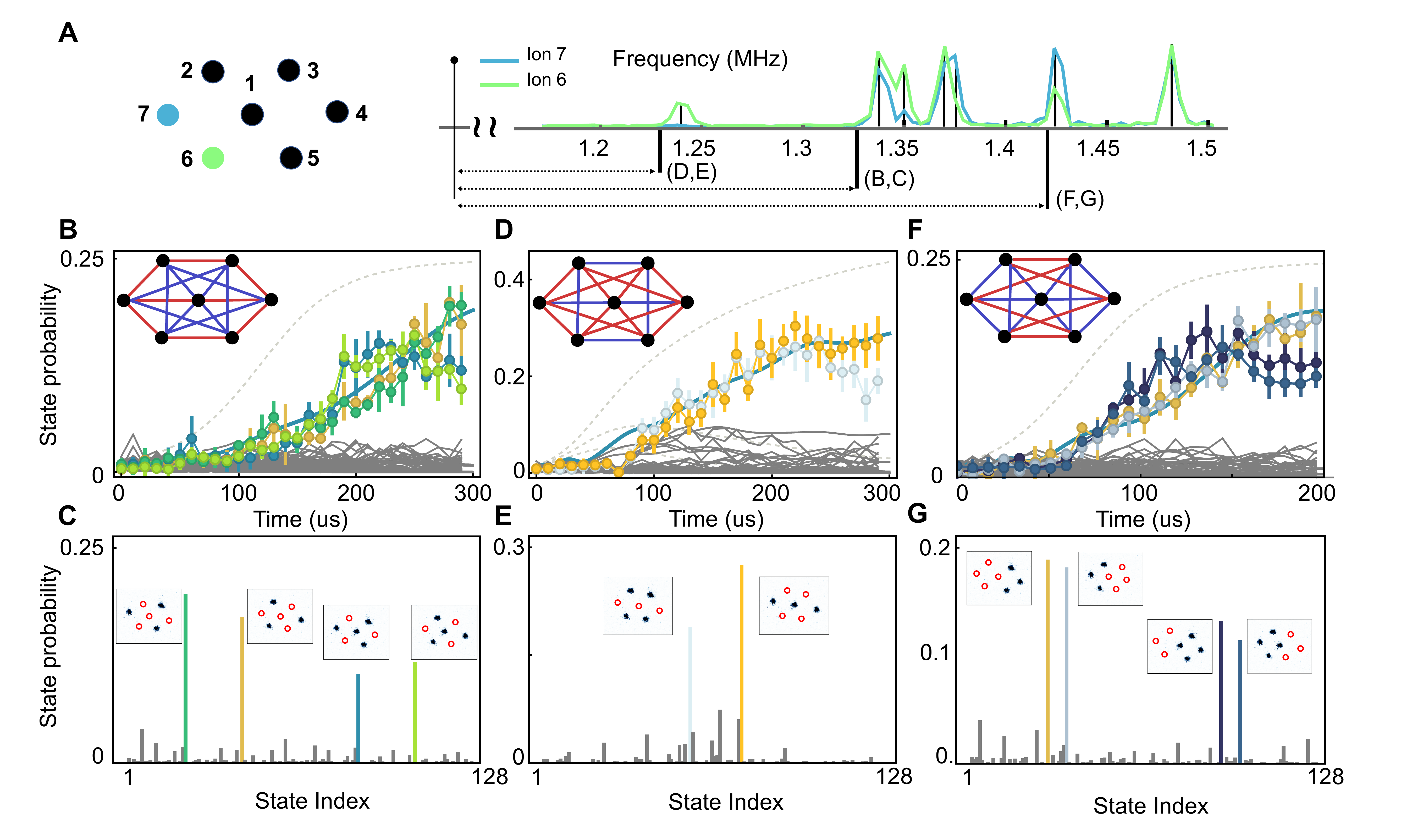}}
\caption{\label{fig:7-ion} {\bf Quantum simulation with a 7-ion 2D crystal.} {\bf a}, Vibrational spectrum of the crystal. The red and green curves represent the fluorescence of red and green ions, respectively. The vertical red lines indicate the mode frequencies, while the solid line represents the detuning used in the experiment. {\bf b},{\bf d},{\bf f} depict the time evolution of different Ising interactions, which are tuned by detunings of the Raman laser beams shown in {\bf a}. The insets represent the interaction diagrams. The points, solid curves, and dashed curves represent the experimental data, theoretical evolution expected from the actual ramp, and the populations in the exact ground state, respectively. {\bf c},{\bf e},{\bf g} depict the experimentally measured populations of all states in binary order at the end of the ramp, which presents the ground states of the Hamiltonians with the given interaction diagrams. Insets depict reconstructed images based on the binary detection of spin states. {\bf b},{\bf c}, is performed with a detuning of \qty{1.328}{\mega\hertz}; {\bf d},{\bf e} is performed with a detuning of \qty{1.231}{\mega\hertz}; {\bf f},{\bf g} is performed with a detuning of \qty{1.416}{\mega\hertz} to the left of the 7th vibrational mode from the COM mode. The error bars represent standard deviation.}
\end{figure*}

We begin with Doppler cooling, which brings the ion crystal into the Lamb-Dicke regime with a mean phonon number of $\bar n\approx 8$. Subsequently, we apply EIT cooling \cite{qiao_2020_eit,feng_2020_eit} for $500\mu$s and then five cycles of sideband cooling to further cool down the crystal near the ground states of the vibrational modes. We initialize all of the spins into $\ket{\downarrow}$ state using the standard optical pumping technique \cite{Olmschenk_2007_manipulation}. The quantum simulation is initiated with a global $\pi/2$ pulse to simultaneously rotate the spins into the eigenstate of the transverse B-field, where the strength of the B-field is sufficiently strong to dominate over the Ising interactions. Then we perform adiabatic evolution as shown in Fig.1{\bf d}. We slowly decrease the strength of B field while maintaining a constant Ising interaction strength. We use a profile of $1/(1+\alpha t)$ for ramping down, which is motivated by local adiabatic evolution \cite{Roland_2002_quantum}, where the speed of the ramping is proportional to the instantaneous energy gap. This ramping profile is then used to prepare a ground state with a much shorter duration than the exponential ramping profiles \cite{Rajibul_2013_emergence, Kim_2010_quantum}. Finally, we measure the individual spin states of the final spin configuration through the standard site-resolved fluorescence detection \cite{camera_detection} using an electron-multiplying charge coupled devices (EMCCD) camera. 

%\section{Engineering of interaction diagram}

We first verify the adiabatic evolution using a 4-ion 2D crystal with a rhombus geometry. Fig.2{\bf a}, shows the vibrational spectrum of the crystal and the Raman detuning used for the generation of the various spin-spin interactions. As shown Fig.\ref{fig:4-ion}({\bf b},{\bf c}), we effectively generate the ferromagnetic interactions for every pair of ions by setting the detuning $\mu$ to be \qty{10}{\kilo\hertz} larger than the frequency of the center-of-mass (COM) mode. Although the detuning generates an anti-ferromagnetic interaction, we effectively reverse the sign of the Hamiltonian by preparing the highest excited states in Fig.2({\bf b},{\bf c}).  Fig.\ref{fig:4-ion}{\bf b}, shows the time evolution of four spins with ferromagnetic interactions, and Fig.2{\bf c}, shows the populations are dominantly in the expected ground states $\ket{\uparrow\uparrow\uparrow\uparrow}$ and $\ket{\downarrow\downarrow\downarrow\downarrow}$ with a probability of $73.34\%$. Figs.2({\bf d},{\bf e}) show the time evolution and the ground states of the Ising model with the nearest anti-ferromagnetic and next-nearest ferromagnetic interactions. We generate the interaction by using the red side of the third mode. The expected ground state, a superposition of alternating Néel spin orders, is prepared with a probability of $64.97\%$. The experimental data of time evolution shown in Figs.2({\bf b},{\bf d}), deviate from the exact ground states but are in agreement with the expected time evolution. This is caused by a faster ramping speed employed which is larger than that a perfect adiabatic condition requires, constrained mainly by the large heating rates of the COM mode.  

We next perform quantum simulation of the transverse Ising models with seven ions in a centered-hexagonal crystal, whose geometrical configuration and vibrational mode spectrum are shown in Fig.\ref{fig:7-ion}{\bf a}. We engineer the strengths and signs of the spin-spin interactions by tuning the detuning of the bichromatic Raman laser beams. In the first case with a detuning of \qty{1.328}{\mega\hertz}, the outer spins exhibit anti-ferromagnetic nearest-neighbor interaction and ferromagnetic next-nearest-neighbor interaction, while the center spin come with four ferromagnetic interactions and two anti-ferromagnetic interactions, as shown in Figs.\ref{fig:7-ion}({\bf b},{\bf c}). Such an interaction diagram has a similar frustration to that of Fig.1{\bf b}, where the outer spins show an alternating order and the center spin experiences competing interactions between the $\ket{\uparrow}$ and $\ket{\downarrow}$ spins. Therefore, the ground state for this interaction diagram contains the superposition of the $\ket{\uparrow}$ and $\ket{\downarrow}$ states for the center spin. Including the equal superstitions of different alternating orders of outer ions, there are four different spin configurations in the ground state. We observe the ground state after the adiabatic evolution. Fig.\ref{fig:7-ion}{\bf b}, compares experimental and theoretical results during the adiabatic evolution. In agreement with theoretical expectations, four spin configurations are dominantly populated with a probability of $58.5\%$ at the end of the evolution as shown in Fig.\ref{fig:7-ion}{\bf c}, which clearly indicates the frustrated spin states of the Hamiltonian.  

In the second case shown in Figs. \ref{fig:7-ion}({\bf d},{\bf e}), at the detuning of \qty{1.231}{\mega\hertz}, the crystal splits into two sub-lattices, which respectively consist of spin 1-4-7 and 2-3-5-6. The interactions within each sub-lattices are ferromagnetic, and the interactions between different sub-lattices are anti-ferromagnetic. This interaction diagram is free of frustration, and the resulting ground state contains the same spins within the same sub-lattices and opposite spins between different sub-lattices. As shown in Fig \ref{fig:7-ion}({\bf d},{\bf e}), the ground state consists of two configurations with a probability of $46.3\%$. In the ground states, one of the sub-lattices is all spin down while the another is all spin up, which is in agreement with theoretical expectations. 

\begin{figure}[!htb]
\center{\includegraphics[width=0.5\textwidth]{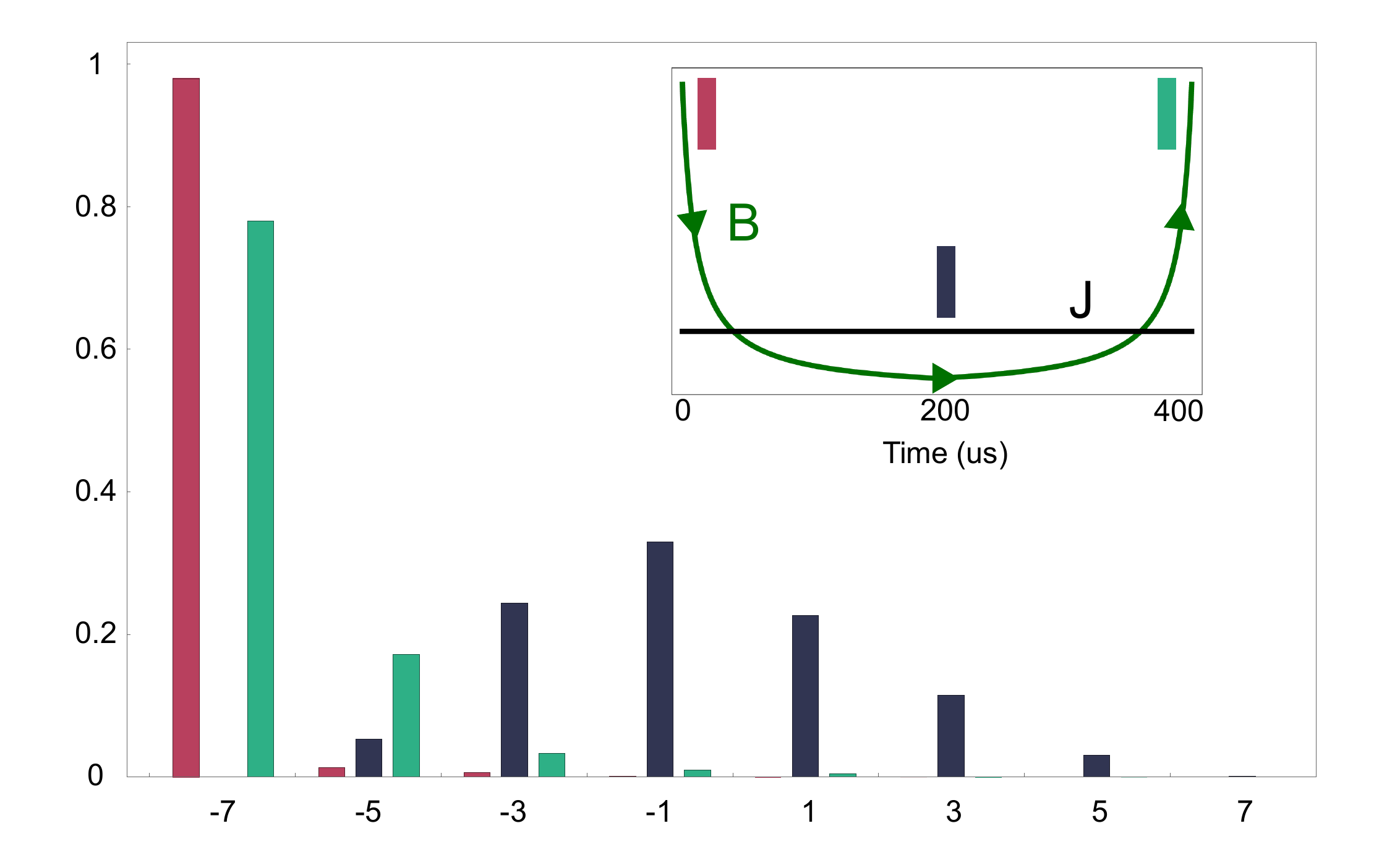}}
\caption{\label{fig:rampingback}{{\bf Quantum coherence probed by time reversal of adiabatic evolution.} Population distributions of each spin-x component in the initial state (red), at the end of adiabatic evolution (black), and after the reversal of adiabatic evolution (green). The initial state is recovered after reversal with fidelity of $80\%$ which strongly indicates the quantum coherence in the simulation. The inset presents the ramping trajectory corresponding to the transverse B field (green curve) and the averaged Ising couplings (black curve). The experimental conditions are the same as in Fig. 2{\bf b},{\bf c}.}
}
\end{figure}

In the third case, the crystal experiences frustration from different type of interactions at the detuning of \qty{1.416}{\mega\hertz}, as shown in Fig.\ref{fig:7-ion}({\bf f},{\bf g}). The crystal can be divided into three parts: a left sub-lattice with spin 2-6-7, a right sub-lattice with spin 3-4-5, and the center spin.  The spins within each sub-lattice have ferromagnetic interactions, while spins between different sub-lattices interact antiferromagnetically. If there is no center spin, the spins in the left sub-lattice have the same orientation, and the spins in the right sub-lattice will be in the opposite orientation without frustration. However, frustration arises from the ferromagnetic interactions between the center spin and all the other spins. Due to the competing interactions on the center spin, the ground state becomes degenerated with four-spin configurations, which are clearly shown in Fig.\ref{fig:7-ion}{\bf f},{\bf g}. The adiabatic evolution prepares the ground state with a probability of $61.45\%$. 

To probe the coherence of the adiabatic evolution we apply a verification scheme by the time-reversed analog simulation, which can sensitively detect incoherent noise \cite{Rajibul_2013_emergence,Shaffer_2021_practical}. The basic experimental protocol first performs adiabatic evolution by ramping down the B field and then reverse the adiabatic evolution by ramping back the B field. Fig 3 shows the measured distribution of the x-component of the total spin operator $\hat{S_{x}}=\sum_{i}\hat{\sigma}_x^{(i)}$. Initially, $90\%$ of the population is prepared in-state $\ket{S_{x}=-7/2}$. The total magnetization at the end of adiabatic evolution is approximately zero since the spins are along the y-direction, which is the direction of the spin-spin interaction. After the time-reversal ramping, $80\%$ of the population returns to the initial state of $\ket{S_{x}=-7/2}$, indicating that the process is coherent. Here, the experimental conditions are the same as that of Fig.3({\bf b},{\bf c}), where the ground state is frustrated. We note that the time-reversal analog verification protocol is insensitive to shot-to-shot parameter fluctuation, parameter miscalibration, and crosstalk \cite{Shaffer_2021_practical}, which would result in insignificant effects in the adiabatic quantum simulation \cite{Shaffer_2021_practical}.

\begin{figure}[!htb]
\center{\includegraphics[width=0.5\textwidth]{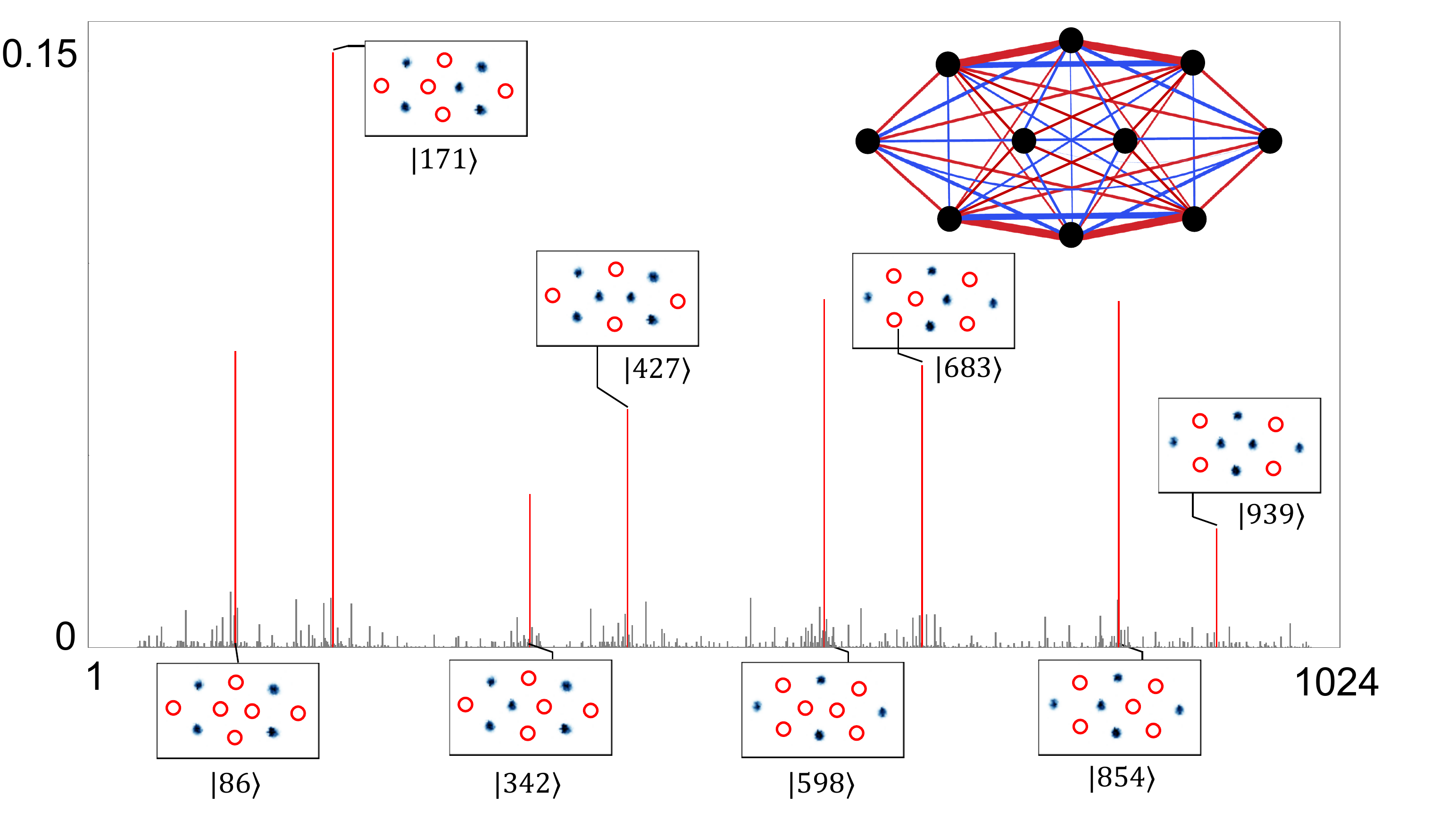}}
\caption{\label{fig:10-ion}{\bf Frustrated quantum magnet of a 10-ion 2D crystal.} The inset depicts the geometry of the spin-spin interaction. The horizontal axis indicates the 1024 computational bases of a 10-spin Hilbert state in binary order based on the numbering of the ions shown in the inset. The vertical axis represents the population in each state. The red bars represent the dominantly populated states with the reconstructed images of the spin configurations.  
}
\end{figure}

We further prepare the ground state of a frustrated Ising model with ten ions, where the geometry and interaction diagram of the ten-ion 2D crystal is shown in Fig. 5. The duration of adiabatic evolution is 300 $\mu$s, which is similar to previous experiments. The experiment is performed at the detuning of \qty{1.296}{\mega\hertz} from the carrier transition, which is located between the 8-th and 9-th modes. The outer eight spins are dominated by the nearest-neighbor anti-ferromagnetic interactions, where the ground state exhibits the Néel order. The two central spins are surrounded by the other eight spins with competing interactions, which results in the degeneration of all possible spin configurations in the ground state. Therefore, the ground state mainly consists of eight configurations of spin states. As clearly shown in Fig \ref{fig:10-ion}, the final probability distribution occupies approximately $50\%$ of the ground-state spin configurations.

In our experiment, the imperfections mainly originate from errors generated while preparing the initial states, heating of the vibrational modes, and non-adiabatic transitions (see Supplementary). The imperfections from initial state preparation are less than 5 $\%$ for the ten-ion crystal and can be further improved using Raman laser beams with larger beam widths. The influence of heating on the spin-spin interactions seriously limits the further scaling up of the number of ion spins for quantum simulation. This can be resolved by enclosing the trap at a cryogenic temperature that significantly reduces the heating rates \cite{Bruzewicz_2015_measurement}. With lower heating, adiabatic evolution can be performed for longer durations and thus suppress the non-adiabatic transitions.

This work reports experimental results of simulating quantum magnets on 2D crystals of ions with the ability to engineer various spin-spin interactions. A variety of programmable interactions are expected to be realized by using various schemes of controlling pulses \cite{Figgatt_2019_parallelgate, lu_2019_globalgate, ozeri_2020_multitone, ozeri_2020_synthetic}, and with this ability our platform can be used to solve classically intractable problems such as quadratic unconstrained binary optimization. Our experimental demonstration can also be implemented to further study 2D quantum spin systems with controllable polynomial-decaying interactions \cite{tj_model_long_range,kuramoto2009dynamics,Haldane_1988,Shastry_1988,Tobias_longrange}. We anticipate that the 2D-ion crystal will be a powerful tool in solving classically intractable problems and be farmland for exploring exotic phenomena that emerged in 2D quantum systems.

\newpage

\bibliography{nature}

\section*{Acknowledgements}
We thank Li You, John Bollinger, and Jim Freericks for carefully reading the manuscript. This work was supported by the National Key Research and Development Program of China under Grants No.\ 2016YFA0301900 and No.\ 2016YFA0301901, the National Natural Science Foundation of China Grants No.\ 92065205, and No.\ 11974200.

\section*{Author information}
\subsection*{Author contributions}
K.K and M.Q conceived the idea and designed the experiments; M.Q., Z.C., Y.W., B.D., N.J. with the assistance of W.C., P.W., C.L., E.G., X.S., and H.T. developed the experimental system; M.Q. with help of J.Z. optimized experimental schemes; M.Q. took and analyzed the data. K.K. supervised the project; and M.Q., K.K., Z.C., E.G., and Y.W. contributed to the writing of the manuscript with the agreement of all the other authors.

\subsection*{Corresponding author}
Correspondence to M.Q, and K.K.

\subsection*{Competing interests}
The authors declare no competing interests.

\section*{Data Availability}
All relevant data are available from the corresponding authors upon request.

\section*{Method}

\subsection*{1: Trap conditions}

To generate 2D ion crystal, we squeeze crystal along y-direction by setting the voltage of electrodes $V_{0}=V_{1}$, $V_2=V_3=V_4=V_5$, and with ratio of $V_{0}/V_{2}=1/5.41$. The ratio of voltages is calculated based on numerical simulation given the trap geometry \cite{wang_2020_coherently}.

For 4-ion and 10-ion experiment, the trap frequencies are $\{\omega_{x},\omega_{y},\omega_{z}\}=\{0.626,0.404,1.503\}$ \unit[mode=text]{\mega\hertz}. For 7-ion experiment, the trap frequencies are $\{\omega_{x},\omega_{y},\omega_{z}\}=\{0.486,0.407,1.482\}$ \unit[mode=text]{\mega\hertz}. The axes of the trap frequencies are shown in Fig. S1. The 2D crystal is in the xy plane, where the x-axis is confined by RF power and the y-axis is controlled by DC-voltage \cite{wang_2020_coherently}. The trap frequencies are the same for 4-ion and 10-ion cases, but for 7-ion case the trap frequencies along the x- and y-direction are closer. With 7 ions, we release the confinement along x-direction to make a hexagonal geometry that 6 ions form a hexagon and 1 ion locates at the center of the hexagon. If we use the setting of the 4-ion case with 7 ions, ions will form a shape of a ladder as shown in Fig S1, which is different from the centered hexagonal geometry. For 4-ion and 10-ion cases, we increase the z-direction trap frequency as high as possible to reduce heating from environment noise meanwhile maintain crystal are 2D and inter-ion distance is around \qty[mode=text]{5}{\micro\meter}. 

\begin{figure}[!htb]
\center{\includegraphics[width=0.4\textwidth]{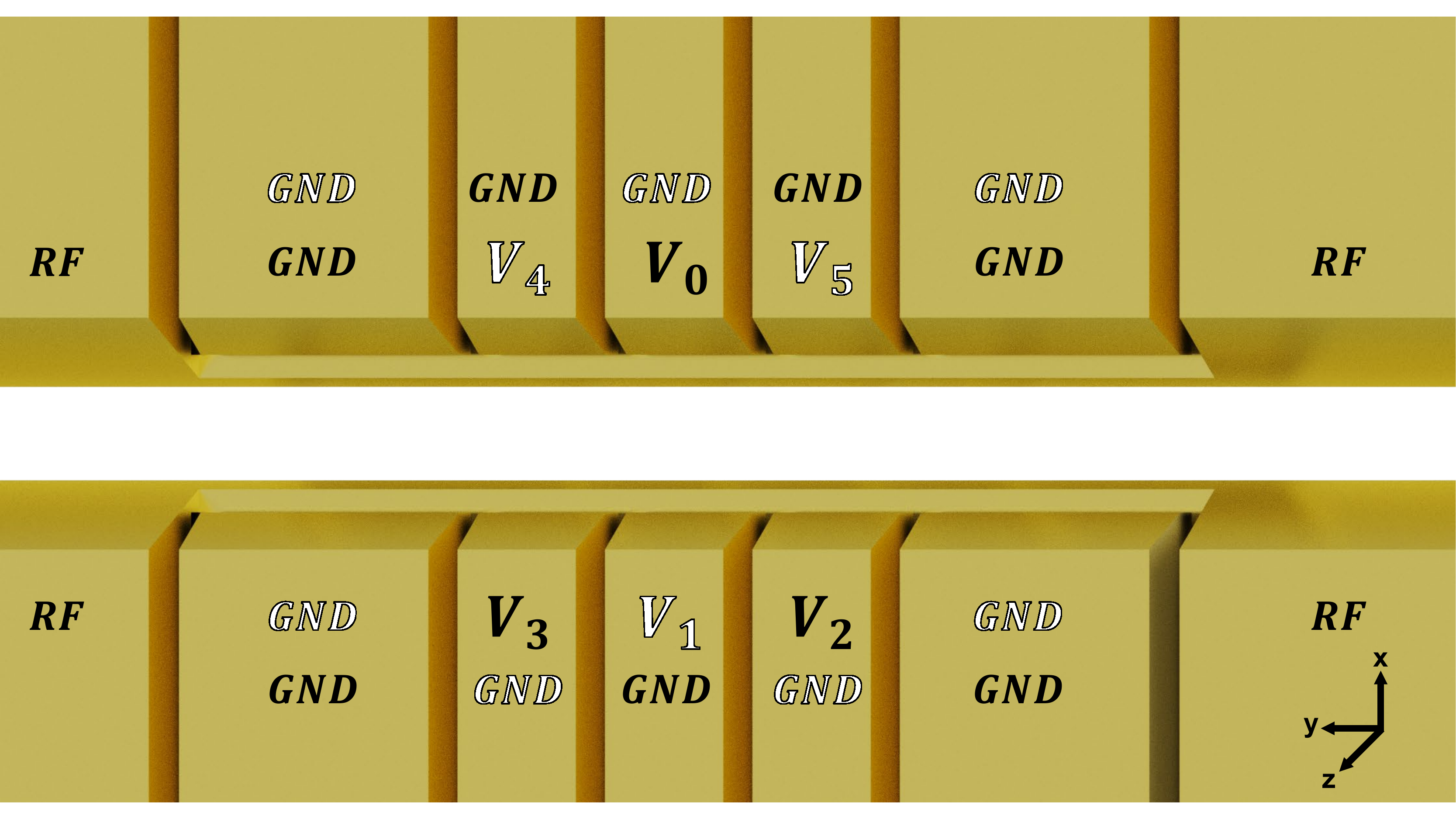}}
\caption{\label{fig:trap_geometry} Trap geometry. Here black texts represent voltage labels of front side of the trap, and white texts represents voltage labels the back side electrodes. The GND electrodes are connected to the ground.}
\end{figure}

\subsection*{2: Compensation of micromotion}

Different from the 1D ion chain, the excess micromotion \cite{berkeland_1998_minimization} of the 2D ion crystal can not be perfectly compensated. Micromotion is the ion-motion synchronous with the oscillating RF field, which degrades the precise control of ions. In the Paul trap, the zero RF field can exist in a line with which we can overlap the chain of ions, which minimizes the effect of micromotion. However, it is not possible for 2D ion crystals because there exists no plane in which RF electric field strength is zero. 

Although the perfect micromotion compensation is impossible, we can still eliminate the effect of micromotion on quantum operations. When the 2D-ion crystal is formed in a plane with only an in-plane electric field, the micromotion exists also only inside the plane. Therefore, we minimize the effect of micromotion on the Raman laser beams by matching the plane of the 2D ion crystal to the plane with only the in-plane electric field and making the net-propagation vector of Raman beams perpendicular to the plane. In our case, the trap electrodes geometry has three symmetric planes, the x-y plane, the y-z plane, and the z-x plane. The RF field is also symmetric under mirror flip for these planes and the symmetry guarantee that in each of the three planes the electric field has only in-plane components. We make the crystal overlapped with the x-y plane by rotating the principal axis of the electric field \cite{wang_2020_coherently}.

\begin{figure}[!htb]
\center{\includegraphics[width=0.5\textwidth]{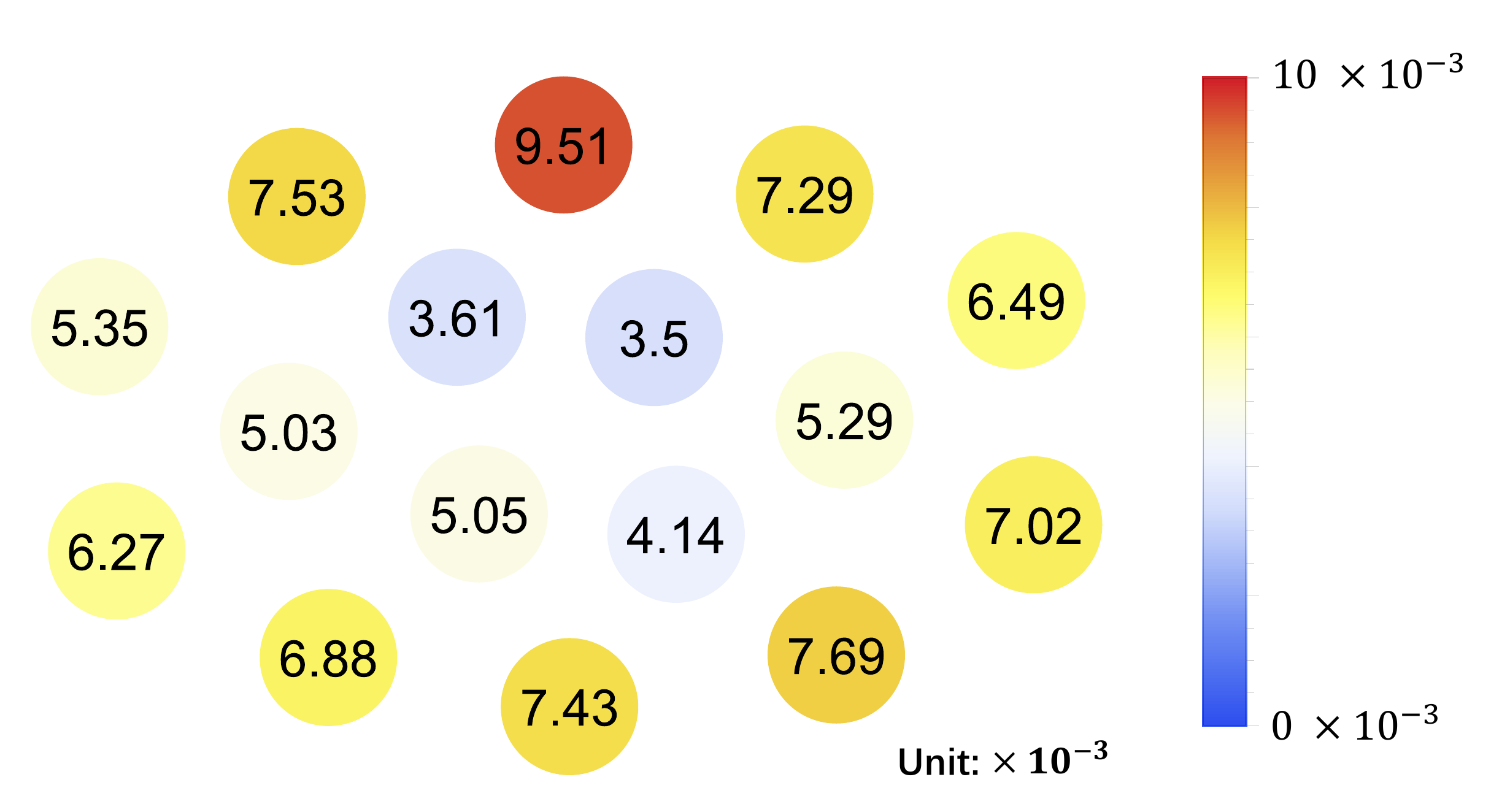}}
\caption{\label{fig:16-ion-micromotion} The relative strengths of  micromotion sidebands with respect to those of carrier transitions for 16 ions.
}
\end{figure}

In the experiment, we match the plane of the 2D-ion crystal to that of an in-plane electric field by using the following three steps, a single ion, a linear chain, and a 2D crystal. With a single ion, we bring the ion to the RF null position by using the traditional method \cite{berkeland_1998_minimization}. Second, we find the condition of electric voltages for the 2D ion crystal and then increase the RF power to change the 2D crystal to a linear chain. We make the chain of ions overlap with the RF null line. We can detect the mismatch between the ion chain and the null line of the RF field by observing the strengths of micromotion sideband transitions of individual ions. We simultaneously vary the voltages $\Delta V_2=-\Delta V_3=\Delta V_4=-\Delta V_5$, shown in Fig. S1 to minimize the mismatch. Finally, we decrease the power of the RF field and recover the 2D ion crystal. After the first two steps, we only need to rotate the plane of the 2D ion crystal about the null line. We simultaneously adjust the voltages of electrodes $\Delta V_{0}=\Delta V_{1}$ and minimize the micromotion sidebands of individual ions. Figure S\ref{fig:16-ion-micromotion} shows the strengths of micromotion sidebands of 16 individual ions in the 2D-ion crystal after minimizing the strengths of micromotion. The strengths of micromotion sidebands are around 200 times smaller than those of carrier transitions on average.  

\subsection*{3: Detection of multi-ion states}

To detect the multi-ion states, we spatially resolve photons emitted by ions for a duration of applying a detection laser. The 2D crystals are imaged by an objective lens oriented perpendicular to the crystals and we use an electron-multiplying charge-coupled device (EMCCD) camera (Andor iXon 897) to detect photons. To increase the signal-to-noise ratio, we fit the distribution of photon counts over pixels with the Gaussian function and use the fitted amplitude as effective photon counts for single-shot measurement.

In experiment, we use a \qty[mode=text]{1.0}{\milli\second} detection pulse, and we collect $34.7$ photons on PMT with a $0.37$ NA objective lens for the bright state of a single ion. In our case, the average detection fidelities for a 4-ion, 7-ion, and 10-ion 2D crystal are $98.2\%$, $97.8\%$, and $98.0\%$.

\subsection*{4: Generating transverse-field Ising interaction}

We generate the Ising interaction by globally driving ions with two \SI{355}{\nano\meter} pico-second pulsed laser beams. These beams have beatnote frequencies $\nu_{\rm{Qubit}}\pm\mu$ that induce a spin-dependent dipole force (SDF). The wavevector difference $\Delta k=2\pi\sqrt{2}/\lambda$ is aligned along the transverse direction of the 2D ion crystal. The interaction diagram can be engineered by varying the beatnote frequencies close to different modes, similar to the linear chain \cite{kim_2009_entanglement, Monroe2021Programmable}.

The transverse B field is effectively generated by addressing Raman transitions resonant to the hyperfine splitting, which is the carrier transition. The phase of Raman beams determines the direction of the B field on the Bloch sphere. Since the SDF has a $\pi/2$ phase difference with the laser field, we use the same laser phase for both carrier transition and the SDF to make the direction of the transverse B field perpendicular to the direction of SDF. 

We can write the Hamiltonian of a single ion interacting with laser field as:
\begin{align*}
    \hat{H}=\sum_{i,j,k}\Omega_{i} \left\{\hat{\sigma}_{+}^{(j)}\left[\hat{I}-i\eta_{k}(\hat{a}_{k}e^{-i\nu_{k} t}+\hat{a}_{k}^\dagger e^{i\nu_{k} t})\right]e^{-i\omega_i t +i\phi_i}\right. 
    \\\left.+\hat{\sigma}_{-}^{(j)}\left[\hat{I}+i\eta_{k}(\hat{a}_{k}e^{-i\nu_{k} t}+\hat{a}_{k}^\dagger e^{i\nu_{k} t})\right]e^{i\omega_i t -i\phi_i}\right\},
\end{align*},
where only the first-order of Lamb-Dicke parameter are reserved. For carrier transition, we drive only one transition, and the Raman detuning matches the hyperfine splitting of qubit. the Hamiltonian reduce to
\begin{equation}
\label{Hamiltonian: carrier}
    \hat{H}=\sum_{j}\Omega_{0} \left(\hat{\sigma}_{+}^{(j)}e^{i\phi}+\hat{\sigma}_{-}^{(j)}e^{-i\phi}\right)=\sum_{j}\Omega_{0} \hat{\sigma}^{(j)}_{\phi},
\end{equation}
which is a spin rotation along $\phi$-axis. $\phi$ is the phase of laser field, and $\Omega_0$ is the Rabi frequency which represents laser field strength.

For SDF, we use beatnote Raman beams, and the Hamiltonian becomes 
\begin{align*}
    \hat{H}&=\sum_{j,k}\Omega_{0} \left[-i\eta_{k}\hat{\sigma}_{+}^{(j)}e^{i\phi}(\hat{a}_{k}e^{-i(\nu_{k}-\omega) t}+\hat{a}_{k}^\dagger e^{i(\nu_{k}-\omega) t})\right.\\
    &\left.+i\eta_{k}\hat{\sigma}_{-}^{(j)}e^{-i\phi}(\hat{a}_{k}e^{-i(\nu_{k}-\omega) t}+\hat{a}_{k}^\dagger e^{i(\nu_{k}-\omega) t})\right]\\
    &=\sum_{j,k}\eta_{k}\Omega_{0} (\hat{a}_{k}e^{-i(\nu_{k}-\omega) t}+\hat{a}_{k}^\dagger e^{i(\nu_{k}-\omega) t})\\
    &\times\left[\hat{\sigma}_{+}^{(j)}e^{i(\phi-\pi/2)}+\hat{\sigma}_{-}^{(j)}e^{-i(\phi-\pi/2)}\right],
\end{align*}
where the last term in the above equation is a spin operator along $\phi-\pi/2$ direction, and the Hamiltonian can be written as
\begin{equation}
\label{Hamiltonian: SDF}
    \hat{H}=\sum_{j,k}\eta_{k}\Omega_{0} (\hat{a}_{k}e^{-i(\nu_{k}-\omega) t}+\hat{a}_{k}^\dagger e^{i(\nu_{k}-\omega) t})\hat{\sigma}^{(j)}_{\phi-\pi/2}.
\end{equation}

Comparing the Eq.(\ref{Hamiltonian: carrier}) and Eq.(\ref{Hamiltonian: SDF}), we can conclude that driving the SDF and the carrier transition naturally have a $-\pi/2$ phase difference if driven by laser fields with the same phase.

\subsection*{5: Experimental methods of finding optimal Ising interactions}

Experimentally, we scan the detuning of SDF with a fixed duration of the adiabatic evolution and find a detuning with the maximum of the ground population after adiabatic evolution. We choose the detuning of \qty{10}{\kilo\hertz} from the mode that shows the largest population of all up and down states. Since the detuning of SDF determines the strengths of Ising interactions, the detuning scan can reveal the performance of the adiabatic evolution by the measured ground state population after adiabatic evolution. The maximum ground-state population indicates optimal Ising interaction, given adiabatic passage of the transverse field with the initial and final strengths of the transverse field at a fixed duration.

\subsection*{6: Error Budget}

\subsubsection*{Imperfection of initial state preparations}
\begin{figure}[!htb]
\center{\includegraphics[width=0.5\textwidth]{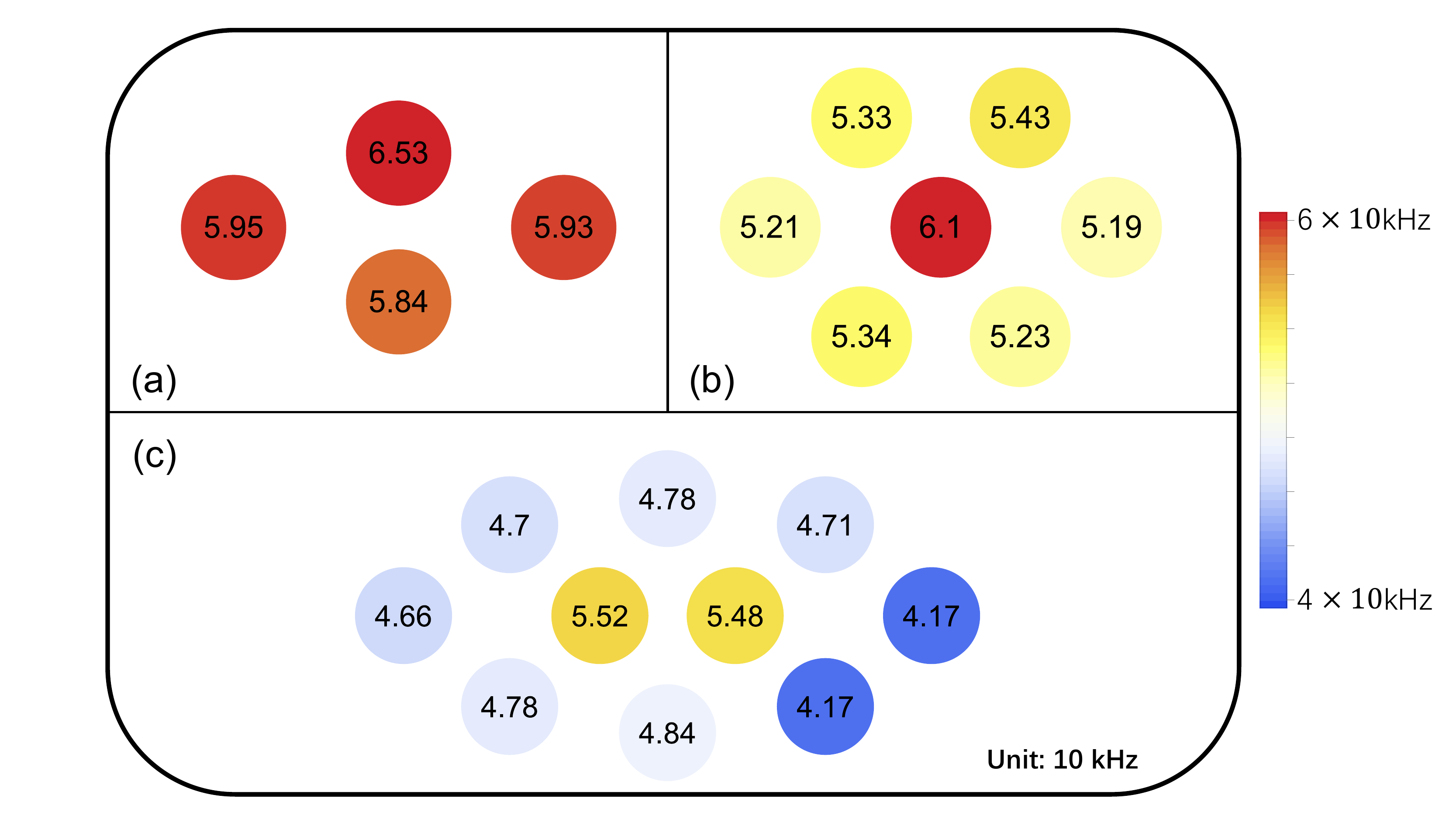}}
\caption{\label{fig:Rabi-distribution} Rabi frequencies (in the unit of 10 kHz) of ions for the 2D crystal with (a) 4, (b) 7, and (c) 10 ions. The deeper color indicates higher Rabi frequency.}
\end{figure}

Limited by the ion-electrode distance, we use Raman beams with a diameter of 25 \unit{\micro\meter}. We observed a serious charging effect as the power and beam size of the Raman beam increased. The relatively small width of Raman beams introduces non-uniform Rabi frequencies over ions, which leads to errors in the global $\pi/2$ rotation. The individual Rabi frequencies depending on the total number of ions are shown in Fig.~\ref{fig:Rabi-distribution}.

The non-uniform Rabi frequencies introduce imperfections in initial state preparation. Given experimentally measured Rabi frequencies, we theoretically estimate the fidelities of initial states for 4-ion, 7-ion, and 10-ion crystals as $99.5\%$, $98.7\%$, and $95.2\%$, respectively. In the experiment, we estimate the fidelities by using Bhattacharyya distances \cite{Bhattacharyya_1946_on} between the measured and ideal populations of initial states, which is equal to state fidelity in our case. 
For 4, 7, and 10-ion crystals, Bhattacharyya distances are $99.2\pm 0.7\%$, $96.3\pm 0.6\%$,  and $83\%\pm 2\%$ respectively, which are consistent to those of theoretical estimations except 10-ion case. The large deviation for 10-ion crystal mainly comes from the insufficient number of measurements for the multi-ion states. Therefore, we use the product of all Bhattacharyya distances of single-ion distribution, which should be also same to the state fidelities for our initial states. The formula is written as 
$$\prod_{i=1}^{N}\left(\sqrt{\frac{1}{2}}\sqrt{p_{i,1}}+\sqrt{\frac{1}{2}}\sqrt{1-p_{i,1}}\right),$$
where $p_{i,1}$ is the upper state population of the i-th ion, and $N$ is the number of ion. For 4, 7-ion, and 10-ion crystals, the experimentally measured Bhattacharyya distances of single-ion distribution are $99.7\%\pm0.2\%$, and $99.6\%\pm0.7\%$, and $95\%\pm5\%$, respectively.
\subsubsection*{Errors from heating of the center of mass modes}

We study the effect of heating of the center of mass modes by using numerical simulation, in particular, with four ions. We use the detuning of Raman beams 10~kHZ that produces all-to-all ferromagnetic interaction as shown in Fig. 2(b) in the main text. 
\begin{figure}[!htb]
\center{\includegraphics[width=0.5\textwidth]{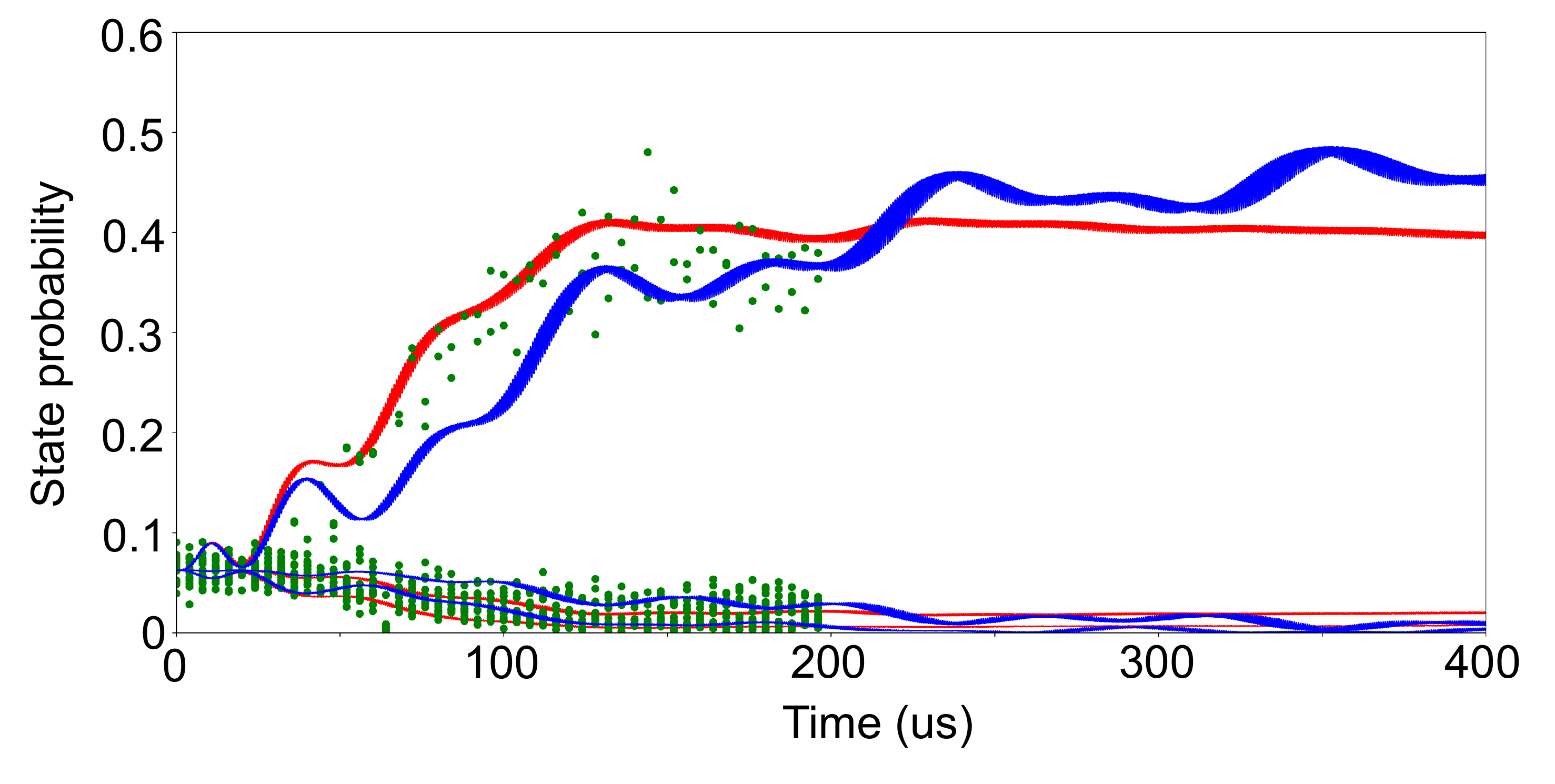}}
\caption{\label{fig:four-ion-heating}  Effect of vibrational heating. The final ground states are the ferromagnetic states with  $\ket{\uparrow\uparrow\uparrow\uparrow}$ and $\ket{\downarrow\downarrow\downarrow\downarrow}$. Here blue curves represent an adiabatic evolution without heating and the rd curves represent an adiabatic evolution with heating of 3200 quanta/s. Green dots represent the experiment result. With heating, the adiabatic evolution results in a lower population of the ground state than that of the adiabatic evolution without heating. We extend the duration to 400 $\mu$s for the numerical simulation, which clearly reveals the limitation of heating. 
}
\end{figure}

In our experiment, the heating rate of a single ion is around $800$ quanta/s, which linearly increases with the number of ions for the center of mass mode \cite{qiao_2020_eit,feng_2020_eit}. For the other modes, it has been known that the heating rates are much smaller than that of the center of mass mode \cite{qiao_2020_eit,feng_2020_eit}. We estimate the imperfections of the final ground state after the adiabatic evolution from the heating of the center of mass mode by numerical simulation.

We include heating by introduce two Lindblad operators $\alpha \hat{a}^\dagger, \alpha \hat{a}$ in master equation.

\begin{equation}
    \frac{d\hat{\rho}}{dt}=-\frac{i}{\hbar}[\hat{H},\hat{\rho}]+\mathcal{L}(\alpha\hat{a})+\mathcal{L}(\alpha\hat{a}^\dagger),
\end{equation}
where $\mathcal{L}(A)=\hat{A}\hat{\rho}\hat{A}^{\dagger}-\frac{1}{2}\hat{A}^{\dagger}\hat{A}\hat{\rho}-\frac{1}{2}\hat{\rho}\hat{A}^{\dagger}\hat{A}$.

To get the value of $\alpha$ consistent with experiment, we can first let $\hat{H}=\hbar\hat{a}^\dagger\hat{a}$ and $\hat{\rho}=\sum_{n=0}^{\infty}c_{n}\ket{n}\bra{n}$. We can get a system of differential equations

\begin{equation}
\left\{
\begin{aligned}
    &\dot{c}_{n}=\alpha^2\left[(n+1)c_{n+1}-(2n+1)c_{n}+nc_{n-1}\right],\quad \text{for } n>0\\
    &\dot{c}_{0}=\alpha^2(c_{1}- c_{0})
\end{aligned}\right.
\end{equation}
Substituting Eq.(4) into the time derivative of average phonon number $\dot{\bar n}=\sum_{n=0}^{\infty}n\dot{c}_{n}$, we can get

\begin{equation}
\label{eq:heating}
    \dot{\bar{n}}=\alpha^2\to \bar{n}=n_{0}+\alpha^2 t,
\end{equation}which indicates a linearly increasing average phonon number. Eq.(\ref{eq:heating}) shows that the value of $\alpha$ is related to the time unit used in simulation. In simulation, we use \unit{\micro\second} as time unit, and expect the average phonon number increase $0.8$ quanta after 1000 \unit{\micro\second} evolution. Since $\Delta\bar{n}=\dot{\bar{n}}\Delta t$, we can calculate the value of $\alpha$ should be $0.028$ for a heating rate of $8\times10^{-4}$ quanta/\unit{\micro\second}. 

In simulation of adiabatic evolution, we use a Hamiltonian with one mode and several spins. 
\begin{equation}
    \hat{H}=\sum_{i}\Omega_{i}\sigma_{x}^{(i)}(\hat{a}e^{-i\nu t}+\hat{a}^\dagger e^{i\nu t})
\end{equation}

Our numerical simulations show that the effect of heating can reduce the population of the ground state at the end of adiabatic evolution. At the optimized detuning, the adiabatic evolution is estimated to generate the population of $96\%$ in the ferromagnetic states for the case of no heating. Including a heating rate of $3200 $ quanta/s on the center-of-mass mode (for four ions), the population of ferromagnetic states drops to $80\%$, which can explain the experimental result of $73\%\pm 5\%$, as shown in Fig S\ref{fig:four-ion-heating}. The further difference can come from the other experimental imperfections such as the dephasing of vibrational modes, intensity fluctuation of lasers, and so on.

\end{document}